\documentstyle[sprocl,epsfig]{article}

\def\Journal#1#2#3#4{{#1} {\bf #2}, #3 (#4)}

\def\PRL{\em Phys. Rev. Lett.}

\def\RMP{\em Rev. Mod. Phys.}
\def\PTP{\em Progr. Theor. Phys.}

\def\be{\begin{equation}}
\def\ee{\end{equation}}
\def\bea{\begin{eqnarray}}
\def\eea{\end{eqnarray}}
\def\pinn{$K^{+} \rightarrow \pi^{+} \nu \bar{\nu}$ }

\begin{document}

\title{STATUS OF THE STUDY OF THE RARE DECAY\\ 
       ${\bf K^{+} \rightarrow \pi^{+} \nu \bar{\nu}}$ AT BNL}

\author{TAKESHI\  K.\ KOMATSUBARA\\
        (for the E787 and E949 collaborations)}

\address{High Energy Accelerator Research Organization(KEK), Tanashi branch\\
         Midori-cho 3-2-1, Tanashi-shi, Tokyo 188-8501, Japan\\
         E-mail: takeshi.komatsubara@kek.jp}

%%%%%%%%%%%%%%%%%%%%%%%%%%%%%%%%%%%%%%%%%%%%%%%%%%%%%%%%%%%%%%
% You may repeat \author \address as often as necessary      %
%%%%%%%%%%%%%%%%%%%%%%%%%%%%%%%%%%%%%%%%%%%%%%%%%%%%%%%%%%%%%%

\maketitle\abstracts{ The current status of the experimental study 
of the rare kaon decay \pinn at Brookhaven National Laboratory (BNL)
by the E787 and E949 collaborations is reported.}

\section{Theoretical Motivation}

The \pinn decay is a flavor changing neutral current process
induced in the Standard Model (SM) by loop effects 
in the form of penguin and box diagrams. 
The decay is sensitive to top-quark effects 
and provides an excellent route to determine the absolute value of
$V_{td}$ in the Cabibbo-Kobayashi-Maskawa matrix.
Long-distance contributions are negligible  
and the hadronic matrix element is extracted from the 
$K^+\to\pi^0e^+\nu$ decay.
The theoretical uncertainty is 7\% 
from the charm-quark contribution in the next-to-leading-logarithmic 
QCD calculations~\cite{BBL}.

The branching ratio is represented in the SM as~\footnote{
   $X(x_{t})$ is the Inami-Lim loop function~\cite{IL}
   with the QCD correction,  
   $x_t \equiv m_t^2/m_W^2$, and 
   $\rho_0$ is due to the charm contribution.} :
\begin{equation}
 B(K^+\to\pi^+\nu\bar{\nu}) = 
 4.57 \times 10^{-11} \times A^{4}\times X(x_{t})^{2}
 \times [\ (\rho_{0}- \rho )^{2}\ +\ \eta^{2}\ ]
\end{equation}
in the Wolfenstein parameterization $A$, $\rho$ and $\eta$. 
With the $\rho$-$\eta$ constraints from other K and B decay
experiments, the SM prediction of the branching ratio~\cite{BBL} 
is $(0.6-1.5)\times10^{-10}$.
New physics beyond the SM
could affect the branching ratio.
In addition, the two-body decay $K^+\to\pi^+X^0$, 
where the $X^0$ is a weakly-interacting light particle  
such as a familon~\cite{familon}, 
could also be observed as a '$\pi^+$ plus nothing' decay. 
Since the effects of new physics are not expected to be too large, 
a precise measurement of a decay at the level of $10^{-10}$ is required.

\section{E787 Detector and Analysis}

Experiment 787~\footnote{
   E787 is a collaboration of BNL, Fukui, KEK, 
   Osaka, Princeton, TRIUMF and Alberta.}
at the Alternating Gradient Synchrotron (AGS)
of BNL
performed an initial search in 1989-91 
and obtained the 90\% confidence level upper limit~\cite{E787-0} 
of $2.4 \times10^{-9}$.
Following major upgrades of the detector and the beam line, 
E787 took data from 1995 to 1998.

\begin{figure}[t]
 \leavevmode
 \hfil
 \epsfig{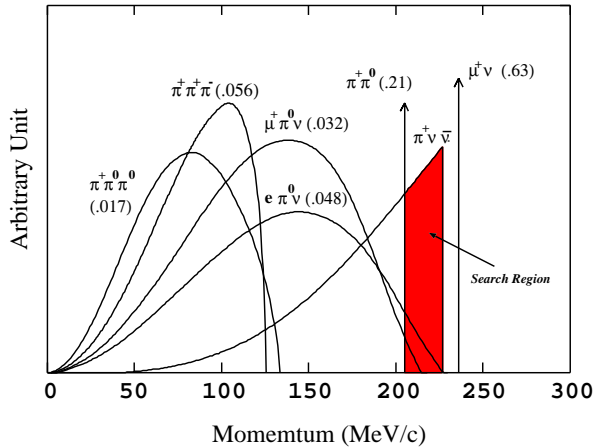}
 \hfil
\caption{Momentum spectrum of the charged particles
 from stopped $K^+$ decays. \label{fig:spctl}}
\end{figure}

E787 measures the charged track emanating from stopped $K^+$ decays. 
The $\pi^+$ momentum from \pinn is 
less than 227MeV/$c$ as shown in Figure~\ref{fig:spctl}, 
while the major background sources of 
$K^+\to\pi^+\pi^0$ ($K_{\pi 2}$, 21.2\%) and 
$K^+\to\mu^+\nu$   ($K_{\mu 2}$, 63.5\%)
are two-body decays and have monochromatic momentum of 
205MeV/$c$ and 236MeV/$c$, respectively.
The region ``above the $K_{\pi 2}$'' between 211MeV/$c$ and 230MeV/$c$ 
is adopted for the search.  

\begin{figure}[t]
 \leavevmode
 \hfil
 \epsfig{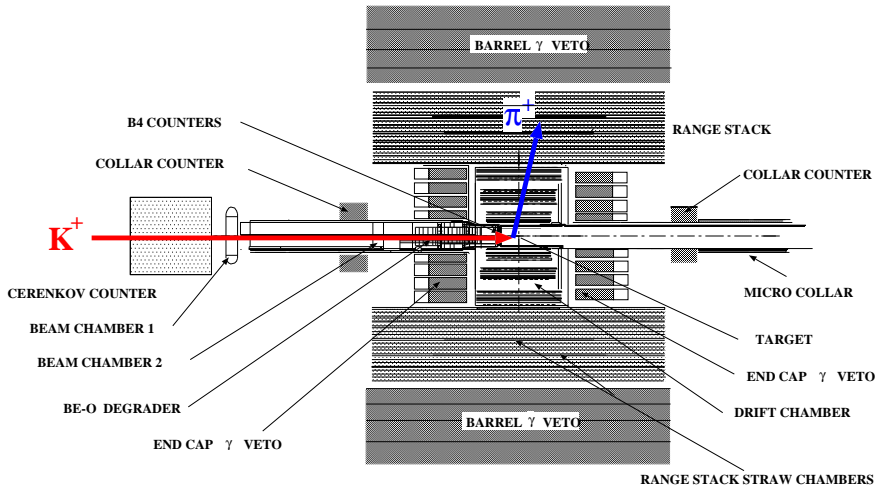}
 \hfil
\caption{Schematic side view of the E787 detector. 
   \label{fig:detector}}
\end{figure}

The E787 detector (Figure~\ref{fig:detector})  
is a solenoidal spectrometer
with the 1.0 Tesla field directed along the beam line. 
Slowed by a BeO degrader, kaons stop in the scintillating-fiber target 
at the center of the detector. 
A delayed coincidence requirement ($>$ 2nsec) of the timing 
between the stopping kaon and the outgoing pion
helps to reject backgrounds of pions scattered into the detector
or kaons decaying in flight.
Charged decay products pass through the drift chamber, 
lose energy by ionization loss and stop 
in the Range Stack made of plastic scintillators and straw chambers.
Momentum, kinetic energy and range are measured 
to reject the backgrounds by kinematic requirements. 
For further rejection of $\mu^+$ tracks, 
the output pulse-shapes of the Range Stack counters
are recorded and analyzed so that the decay chain 
$\pi^+\to\mu^+\to e^+$ is identified in the stopping scintillator. 
$K_{\pi 2}$ and other decay modes with extra particles
($\gamma$, $e$, ...) are vetoed by the in-time signals 
in the hermetic shower counters. 

Extremely effective background suppression
is required in this experiment, 
and reliable estimation of the rejections 
is essential to interpret potential observations. 
Data rather than Monte Carlo    
are used to do background studies.
A set of background samples is prepared
by reversing some of the selection cuts~\footnote{
   For example : the $K_{\pi 2}$ backgrounds are rejected 
   by kinematic cuts and photon veto cuts.  
   By reversing the veto and requiring 
   photons from $\pi^0$ in the detector, 
   the tails in the $\pi^+$ kinematic distributions are studied. 
   By picking up events 
   with the track momentum/energy/range in the $K_{\pi 2}$ peak 
   and by applying the photon veto cuts, 
   the rejection of the veto is checked.}, 
which also assures that 
the development of the cuts and estimates of the background levels
are made without looking at the candidate events 
(``blind'' analysis). 
Furthermore, background studies are performed 
with partial data samples 
and the results are confirmed using the full sample. 
Possible correlations of the cuts are investigated.
The background levels around the signal region are predicted by loosening cuts 
and are confirmed using data. 
Background level shapes 
inside the signal region are calculated in advance
in the form of likelihood functions. 

In the 1995 data set, 
with the total kaons stopping in the target $1.49\times 10^{12}$
and the acceptance $0.16\%$, 
one event, shown in Figure~\ref{fig:event},
was observed~\cite{E787-95} in the signal region. 
The estimated background level ($0.08\pm 0.03$ events) 
corresponds to a branching ratio of 3$\times 10^{-11}$. 
The measured branching ratio is $(4.2^{+9.7}_{-3.5})\times 10^{-10}$ 
($0.006 < |V_{td}| < 0.06$),  
which is consistent with the SM prediction. 

\begin{figure}[t]
 \leavevmode
 \hfil
 \epsfig{file=event.epsi,height=1.85in,angle=0}
 \hfil
\caption{Evidence for the decay \pinn. 
   \label{fig:event}}
\end{figure}

\section{E787 in 1996, 97 and 98}

The total acceptance of 0.16\% includes
the phase space above the $K_{\pi 2}$ (0.16), solid angle acceptance
of the charged track (0.39), $\pi^+$ stop without nuclear interaction 
nor decay-in-flight in the detector (0.50), 
and the acceptance of $\pi^+$ identification with 
the $\pi^+\to\mu^+\to e^+$ decay chain (0.25)
to achieve the $\mu^+$ rejection of $10^{5}$.
The main limitation due to
the requirement of no extra hits in the detector
at the decay time 
is applied in the analysis
costing around 40\% of the acceptance.
The strategy of E787 for the post-1995 runs was therefore 
to limit the instantaneous rates and attempt to gain in the overall number of 
stopped kaons. 

In the experiment, only 20\% of the kaons in the beam 
are slowed down and stop in the target while the remainder
are lost or scattered out in the degrader.
The rates in the E787 detector are proportional
to the incident kaons, not to the stopped kaons in the target. 
That means, with the same incident flux and a lower beam-momentum, 
the kaon stopping fraction increases while accidental hits decrease.
Also, using additional proton intensity to extend the spill length 
without increasing the instantaneous rate raises the number
of kaon decays per hour without impacting the acceptance.

By reducing the beam-momentum from 790MeV/$c$ in 1995
to 710MeV/$c$ in 1997 
the fraction of incident kaons stopping in the target was improved by 44\%.
The AGS spill length was extended from 1.6sec to 
2.2sec in 1998.
Other improvements in the trigger and readout 
provide acceptance gains of about 20\%.
In the off-line analysis,  
kinematic codes with better resolution and rejection-power 
were developed, and 
in the current study of the combined 1995-97 data sets
more rejection corresponding to the background level of $1\times 10^{-11}$ 
has been achieved with minimal loss in acceptance. 
The analysis is ongoing. 
The expected sensitivity for the the entire E787 data should reach 
less than $0.9\times 10^{-10}$.

\section{The New E949 Experiment}

From 1999 the Relativistic Heavy Ion Collider (RHIC) at BNL
starts operation, and 
the AGS is primarily used as the RHIC injector.  However it is
required  for only $\sim 4$hours per day for this purpose.
The rest of the time can be used for the proton program.
The new experiment 949~\footnote{
   E949 is a collaboration of Alberta, BNL, Fukui, INR, KEK, 
   Osaka and TRIUMF. The information is available at 
   \noindent
   {\sf http://www.phy.bnl.gov/e949/}.}
is to continue the experimental study of \pinn 
at the AGS based on the E787 experience. 
An additional photon veto system 
will be installed in the detector to improve the photon rejection. 
E949 aims to reach a sensitivity of $10^{-11}$ or less
in two to three years of operation.

\section{Summary}
E787 has observed the evidence for the \pinn decay in the 1995 data set
and, with the entire data through 1998, 
expects to reach the sensitivity better than $0.9\times 10^{-10}$.
The new E949 continues the study at the BNL-AGS.

\section*{Acknowledgments}
This research was supported in part 
by the U.S. Department of Energy under Contracts 
No. DE-AC02-98CH10886 
and No. W-7405-ENG-36, 
and Grant No. DE-FG02-91ER40671, 
by the Ministry of Education, Science, Sports and Culture of Japan (Monbusho), 
and by the Natural Sciences and Engineering Research Council 
and the National Research Council of Canada. 
The author would like to acknowledge support 
from Grant-in-Aid for Encouragement of Young Scientists 
by Monbusho.


\section*{References}
\begin{thebibliography}{99}
 \bibitem{BBL}G. Buchalla, A.J. Buras and M.E. Lautenbacher,
    \Journal{\RMP}{68}{1125}{1996}.
 \bibitem{IL}T. Inami and C.S. Lim,
    \Journal{\PTP}{65}{297; 1172(E)}{1981}.
 \bibitem{familon}F. Wilczek, \Journal{\PRL}{49}{1549}{1982}.
 \bibitem{E787-0}S. Adler {\it et al}, \Journal{\PRL}{76}{1421}{1996}.
 \bibitem{E787-95}S. Adler {\it et al}, \Journal{\PRL}{79}{2204}{1997}.
\end{thebibliography}
\end{document}